# Chiral charge density waves induced by Ti-doping in 1$T$-TaS$_2$

J. J. Gao[1,2], W. H. Zhang[3], J. G. Si[1,2], X. Luo[1], J. Yan[1,2], Z. Z. Jiang[1,2], W. Wang[1,2], H. Y. Lv[1], P. Tong[1], W. H. Song[1], X. B. Zhu[1], W. J. Lu[1], Y. Yin[3,5], and Y. P. Sun[4,1,5]

[1] *Key Laboratory of Materials Physics, Institute of Solid State Physics, HFIPS, Chinese Academy of Sciences, Hefei, 230031, China*

[2] *Science Island Branch of Graduate School, University of Science and Technology of China, Hefei, 230026, China*

[3] *Zhejiang Province Key Laboratory of Quantum Technology and Device, Department of Physics, Zhejiang University, Hangzhou, 310027, China*

[4] *Anhui Province Key Laboratory of Condensed Matter Physics at Extreme Conditions, High Magnetic Field Laboratory, HFIPS, Chinese Academy of Sciences, Hefei, 230031, China*

[5] *Collaborative Innovation Center of Advanced Microstructures, Nanjing University, Nanjing, 210093, China*

## Abstract

We investigate the Ti-doping effect on the charge density wave (CDW) of 1$T$-TaS$_2$ by combining scanning tunneling microscopy (STM) measurements and first-principle calculations. Although the Ti-doping induced phase evolution seems regular with increasing of the doping concentration ($x$), an unexpected chiral CDW phase is observed in the sample with $x$ = 0.08, in which Ti atoms almost fully occupy the central Ta atoms in the CDW clusters. The emergence of the chiral CDW is proposed to be from the doping-enhanced orbital order. Only when $x$ = 0.08, the possible long-range orbital order can trigger the chiral CDW phase. Compared with other 3$d$-elements doped 1$T$-TaS$_2$, the Ti-doping retains the electronic flat band and the corresponding CDW phase, which is a prerequisite for the emergence of chirality. We expect that introducing elements with a strong orbital character may induce a chiral charge order in a broad class of CDW systems. The present results open up another avenue for further exploring the chiral CDW materials.

Corresponding authors: xluo@issp.ac.cn, wjlu@issp.ac.cn, yiyin@zju.edu.cn and ypsun@issp.ac.cn

# Introduction

Chiral effects are widespread in nature and are of fundamental importance in science, from astrophysics to microscopic molecules [1,2]. A chiral system is an object that cannot be superimposed on its mirror image by any translations or rotations. The chiral effect is also a manifestation of the broken symmetry. Many exciting phenomena occur due to the chiral effect. To explore the chiral origin has always been a fascinating research topic in physics [3,4].

The charge density wave (CDW) and spin density wave (SDW) are two archetypical examples of symmetry breaking in materials, with the order characterized by a complex order parameter [5,6]. The spiral phase is difficult to appear in the CDW phase with a scalar order but could be found in the SDW phase with a vectorial order [7,8,9]. Surprisingly, a definite chiral CDW pattern has been observed in the layered transition metal dichalcogenide (TMD) $1T$-TiSe$_2$ using a low-temperature scanning tunneling microscopy (STM) [10,11]. The simultaneous orbital orders have been proposed to explain the unexpected emergence of the chiral CDW phase. This work then inspired a series of subsequent studies of chiral CDW in TMD materials. The chiral CDW phase was observed in the Cu-intercalated $1T$-TiSe$_2$ samples [12]. The chiral charge order was observed in $2H$-TaS$_2$ using STM at 0.1 K, although later claimed to be a polar state with a mirror symmetry in the lattice of the $2H$ phase [13,14].

As one of the most fascinating TMD materials, $1T$-TaS$_2$ is also a carrier for enriching CDW transitions [15,16,17,18,19]. When cooled below ~ 180 K, the system is in the commensurate CDW (CCDW) state with a $\sqrt{13} \times \sqrt{13}$ reconstruction. The unit cell of a so-called Star of David (SD) is rotated by 13.9° against the lattice, in which 12 Ta atoms move slightly towards the 13th central Ta atom. No chiral order has been detected in $1T$-TaS$_2$ in the literature so far.

In this work, we observe a chiral CDW phase in the Ti-doped $1T$-TaS$_2$ system. For a series of Ti-doped $1T$-Ti$_x$Ta$_{1-x}$S$_2$, we find an abnormal ground state in the sample with a doping amount of $x$ = 0.08. A peculiar SD arrangement is discovered in adjacent domains in the STM image of $x$ = 0.08 samples, different from that in other samples with different dopings. The Fourier-transformed (FT) image indicates that the CDW in adjacent domains has different chiralities. The appropriate element doping induces the chirality in $1T$-Ti$_{0.08}$Ta$_{0.92}$S$_2$, although the parent material has no chiral phase. We relate the chiral CDW phase to the coexistence of charge and orbital orders and the further-enhanced orbital order by Ti-doping.

## RESULTS AND DISCUSSION

In Fig. 1, we summarize the phase diagram of 1$T$-Ti$_x$Ta$_{1-x}$S$_2$ derived from the transport and magnetic measurements as a function of temperature $T$ and content $x$. With the increase of Ti doping, the CCDW state is rapidly suppressed, while the nearly commensurate CDW (NCCDW) transition gradually moves to lower temperature and disappears at $x$ = 0.2. The system also gradually changes from an insulating ground state to a conducting ground state. We find an interesting hidden anomaly in this seemingly regular Ti-doping-induced phase evolution, reflected in the intriguing peak at $x$ = 0.08 in the resistivity ratio curve of $\rho_{2K}/\rho_{360K}$. This phenomenon has been discovered as early as 1975, later inspiring various experiments to study its possible origin and peculiar ground state [20,21,22]. Moreover, the number of 0.08, ~ 1/13, reminds us that SD in TaS$_2$ is also composed of 13 Ta atoms. The anomaly peak location may imply a Ti's particular occupation in SD, e. g., one doped Ti atom per SD. Microscopic measurements may be useful in revealing this anomaly. We then performed detailed STM/STS measurements on a series of samples.

Figures 2(a)-(j) shows the STM images of pure and Ti-doped samples ($x$ = 0, 0.01, 0.05, 0.08 and 0.2) and the corresponding Fourier-transformed (FT) images, while Fig. 2(k) presents the average STS spectrum of these samples. With the measurement taken at 4.5 K, the sample with $x$ = 0 is in the CCDW state. In Fig. 2(a), the STM topography presents a triangle array of SD, with each bright spot representing one SD. For the sample with $x$ = 0.01, the triangle array of SD appears in the SD pattern domains, separated by domain walls [Fig. 2(c)]. With the increase of $x$, the number of domain walls increases and the size of domain decreases [Fig. 2(e), 2(g)]. When $x$ increases to 0.2, the SD array becomes distorted, and the domain walls disappear [Fig. 2(i)]. The STM topographies are consistent with the phase diagram from transport measurement.

From the comparison, the results for the sample with $x$ = 0.08 again caught our attention. For the sample with $x$ = 0.01 or $x$ = 0.05, the lattices of the neighboring domains are oriented in the same direction [Fig. 2(c) and 2(e)], leading to the six bright Bragg peaks of SD lattice in Fig. 2(d) and 2(f). For the sample with $x$ = 0.08, although the triangle SD lattice is still maintained in the domain, the SD lattice directions in adjacent domains are rotated at an angle relative to each other [Fig. 2(g)]. This rotation is consistent with the FT image in Fig. 2(h), in which two sets of SD Bragg peaks appear and rotate about 27.8° from each other.

The emergence of two sets of Bragg peaks implies a broken symmetry in $1T$-Ti$_{0.08}$Ta$_{0.92}$S$_2$. To further explore this phenomenon, we present the FT image of the sample with $x = 0.08$ and mark three wave-vector ($q$) components of the Bragg peaks [Fig. 3(a)]. In Fig. 3(b), we show the corresponding Bragg peaks for each $q$. The three $q$ peaks, called $q_1$, $q_2$, and $q_3$, have the same position but different intensities ($I_{q1}$: $I_{q2}$: $I_{q3}$ = 1:0.96:0.80). Three $q$ vectors with successively increasing intensities are rotated anticlockwise in the FT space. According to the definition of a similar phenomenon in TiSe$_2$ [10], we call this SD lattice the anticlockwise phase (see the quantitative analysis of rest six peaks which are not circled in Fig. 3(a) in the supporting materials). Figure 3(c) shows a 44 nm * 44 nm STM image containing two domains, with the boundary domain wall labeled by a white dotted line. For the two areas framed in two boxes, we display their corresponding FT images in Figs. 3(d) and 3(e). In the two adjacent domains, we find an anticlockwise phase and a clockwise phase, respectively (the quantitative analysis of Bragg peaks can be found in the supporting materials). The two phases are also called to have different chiralities, and the boundary between them is called a chiral domain wall. The two sets of Bragg peaks rotate from each other and jointly contribute the twelve Bragg peaks in the large topography's FT image. We have measured three different samples with $x = 0.08$, all of which yielded the clockwise and anticlockwise phases in adjacent domains (see the details in the supporting materials).

A quantum state where the charge and orbital order coexist has been proposed to explain the chiral CDW state in $1T$-TiSe$_2$ [8,23]. While for $1T$-TaS$_2$, several groups recently suggest an orbital-density-wave (ODW) order, which is mainly contributed by the Ta-$5d_{z^2}$ orbital in the central Ta of each SD [24,25]. The orbital order could be crucial for understanding the chiral CDW. We next quantitatively investigate the orbital character of $1T$-TaS$_2$ and Ti$_{1/13}$Ta$_{12/13}$S$_2$, especially for electronic bands cross the Fermi-level. Three unequal positions exist in one SD, with the central atom called Ta$_0$, the atom in the surrounding ring Ta$_1$, and the atom in the outermost ring Ta$_2$. With a site-dependent calculation, we found that the ground state energy is optimally minimized when the Ti atom occupies the central Ta$_0$ position and the total energy of Ti-doped structure with Ta$_0$-site substitution is 17 ~ 127 meV lower than those with Ta$_1$-site and Ta$_2$-site substitution. The following band calculation of $1T$- Ti$_{1/13}$Ta$_{12/13}$S$_2$ then is obtained with doped Ti occupying the central position.

The calculated orbital character of 1$T$- TaS$_2$ indicates that the Ta-5$d_{z^2}$ orbital makes a dominant contribution to the electronic band, consistent with the previous report [25]. Most notably, the weight of the 5$d_{z^2}$ orbital gradually decreases from the center to the edge in each SD. As shown in Fig. 4(c), this trend illustrates an orbital order in each SD of 1$T$- TaS$_2$. The detailed weight ratio is summarized in the supporting materials. We then implement a similar calculation for the doped 1$T$- Ti$_{1/13}$Ta$_{12/13}$S$_2$. With the central Ta atom replaced by a Ti atom, the central atom's orbital weight increases to almost three times that of the original one. As shown in Fig. 4(d), this change corresponds to a much enhanced orbital order in the SD. Similar to 1$T$-TiSe$_2$, the coexistence of the enhanced orbital order and charge order promotes the appearance of chirality [23]. Since the chiral phase destroy the symmetry and reduce the increased energy caused by the enhanced orbital order, it could maintain the system's stability. The previous energy calculations in 1$T$-TiSe$_2$ also support that the chiral phase can minimize the integrated Ginzburg-Landau free energy [8]. For 1$T$- Ti$_{1/13}$Ta$_{12/13}$S$_2$, each SD is doped with one Ti atom, which ideally may lead to a long-range orbital order. This doping is the most likely condition in which the chiral phase appears to compensate for the high-energy and substantial orbital order. Although the Ti atoms also induce local orbital order in random SDs, no long-range orbital order exists for doping smaller than 1/13. For doping larger than 1/13, Ti atoms occupy multiple positions in SD and weaken the local orbital order. Then we do not observe chiral phases for doping away from 1/13.

We further explore why the particular chiral state of SD appears in Ti-doped samples with $x$ = 0.08. The previous reports show that the chiral phenomenon has not been observed in other 3$d$-element doped 1$T$-TaS$_2$ [26]. For example, the SD lattice has been distorted for $x > 0.05$ in Fe-doped 1$T$-Fe$_x$Ta$_{1-x}$S$_2$. To help understand the origin of chiral state, we calculate and compare the band structures of 1$T$-$M_{1/13}$Ta$_{12/13}$S$_2$ ($M$ stands for 3$d$ element, from Sc to Cu). We display the band structures of 1$T$-TaS$_2$ and Ti$_{1/13}$Ta$_{12/13}$S$_2$ in Fig. 4(a) and (b) and put those of other doping elements in the supporting materials. As shown in Fig. 4(a), a flat band exists along the $\Gamma$-$M$-$K$-$\Gamma$ direction. This band is isolated from the underneath dispersive valence band. The energy gap between the flat band and the dispersive band is generally referred to as the CDW gap ($\Delta_{\mathrm{CDW}}$) [27]. The flat band also corresponds to the sharp peak below Fermi-level in the STS spectrum of undoped 1$T$-TaS$_2$ [as shown in Fig. 2(k)]. When the Ti atom replaces the Ta$_0$, the calculated bands in Fig. 4(b) shows that the flat band shifts above the Fermi level. The flat band shift also seems consistent with the STS spectrum's evolution in Fig. 2(k). The spectrum peak with more weight shifts from below to above the Fermi level with the increasing Ti doping.

The flat band still exists for the Ti doping as large as $x = 0.08$, which is compatible with the retained CDW phase at the large doping. From the fully relaxed structure, we obtain the bond length of Ti-S/Ta-S about 2.450/2.498Å. The energy levels of $3d$ orbitals in a Ti atom are also close to those of $5d$ orbitals of Ta atom [28,29]. The similar bond length and orbital energy could explain the retained flat band and CDW phase in Ti-doped $1T$-TaS$_2$. The CDW phase is the prerequisite for the orbital-order formation and the emergence of the chiral CDW phase. For $1T$-TaS$_2$ doped with other $3d$ elements, both the flat band and the CDW phase are destroyed, and no similar chiral CDW phase is discovered.

## Conclusion

We have systematically investigated the series of $1T$-Ti$_x$Ta$_{1-x}$S$_2$ samples and observed a resistivity anomaly at $x = 0.08$. In the same sample of $1T$-Ti$_{0.08}$Ta$_{0.92}$S$_2$, we discover a chiral CDW phase with STM measurements. The uniform doping of Ti atoms enhances each SD's orbital order, and the long-range orbital order could be formed for the sample with $x = 0.08$. By comparing the band structures of various $3d$ element doped $1T$- $M_{1/13}$Ta$_{12/13}$S$_2$, we find that the flat band and CDW phase are well retained only for Ti-doped samples. Researchers may discover a broad class of materials with the chiral phase, by introducing elements with a stronger orbital character. Our work opens up another avenue for further exploring the chiral CDW materials.

## Acknowledgments

This work was supported by the National Key Research and Development Program under Contracts 2016YFA0300404, 2019YFA0308602, the Fundamental Research Funds for the Central Universities in China, and the National Nature Science Foundation of China under Contracts 11674326, 11874357, 11774351, 11974061 the Joint Funds of the National Natural Science Foundation of China and the Chinese Academy of Sciences' Large-Scale Scientific Facility under Contracts U1832141, U1932217, U2032215 and the Key Research Program of Frontier Sciences, CAS (QYZDB-SSW-SLH015) and The uses with Excellence and Scientific Research Grant of

**Figure 1:**

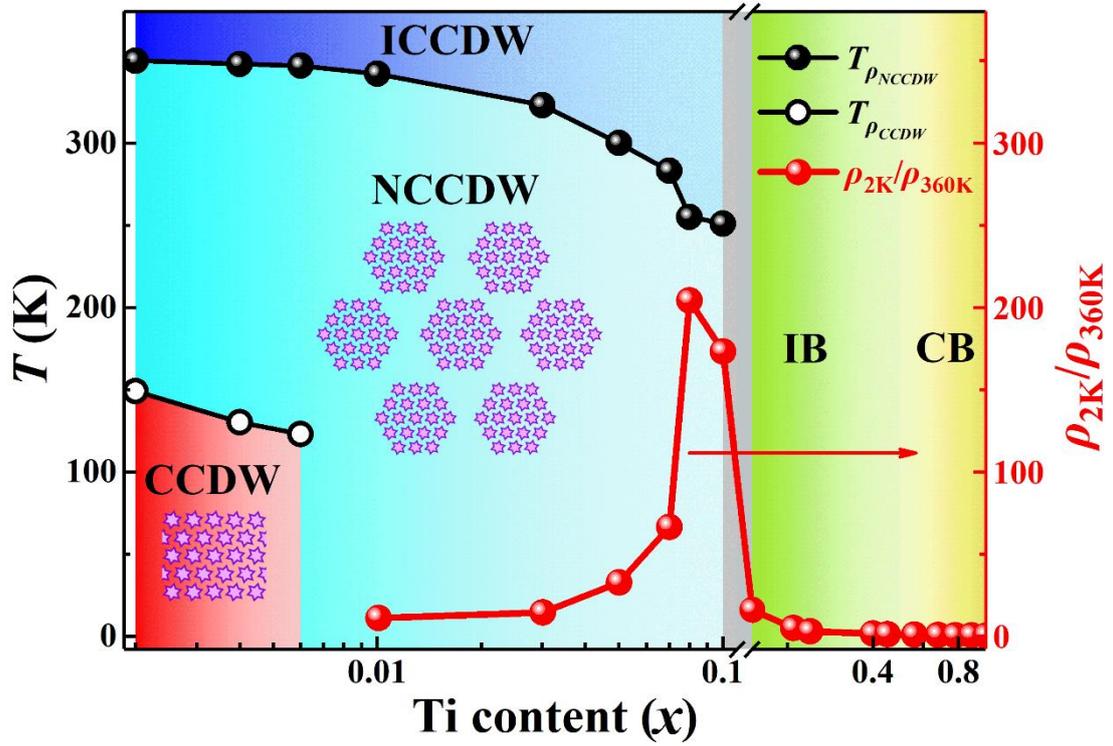

**Fig. 1 (color online):** Phase diagram of $1T$-Ti$_x$Ta$_{1-x}$S$_2$ derived from transport measurement as a function of the temperature $T$ and content $x$, where CCDW, NCCDW, ICCDW, IB, and CB represent the commensurate CDW, nearly commensurate CDW, incommensurate CDW, insulator behavior, and conductor behavior. The ratio $\rho_{2K}/\rho_{360K}$, as a function of Ti concentration $x$, is represented by the red curve in the phase diagram.

**Figure 2:**

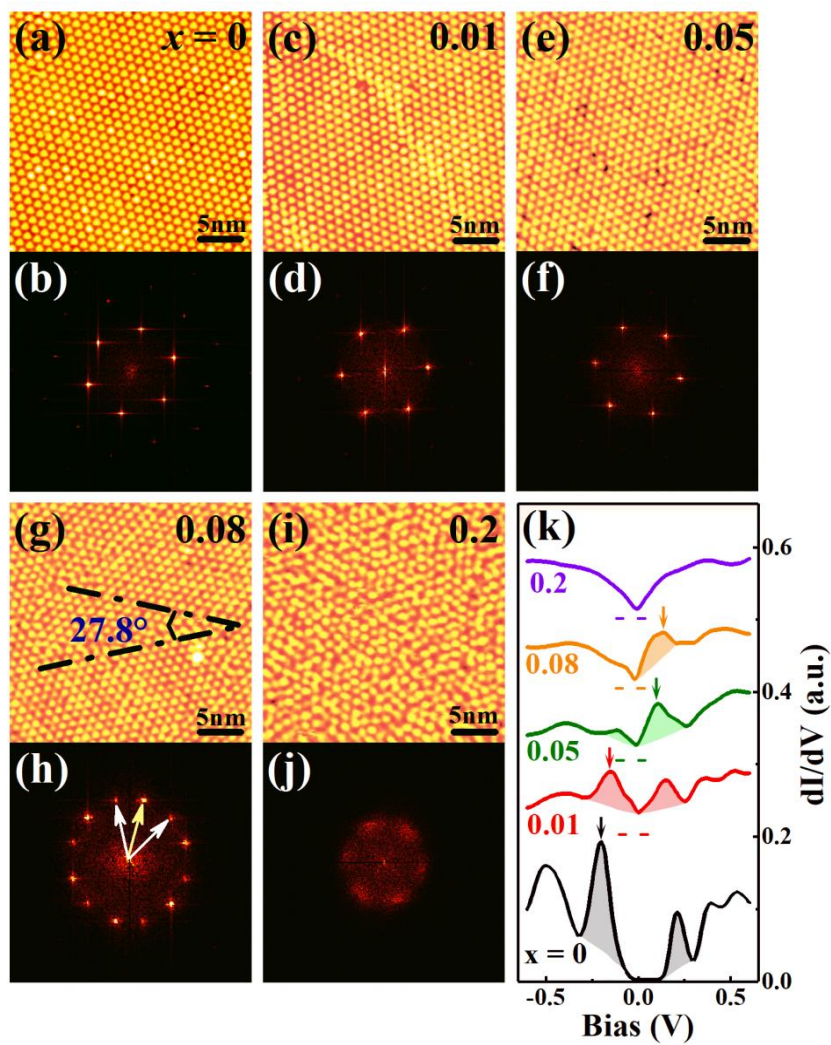

**Fig. 2 (color online):** (a)-(j) STM topographic images and the corresponding FTs taken from single crystal $1T$-$Ti_xTa_{1-x}S_2$ ($x$ = 0, 0.01, 0.05, 0.08 and 0.2) at 4.5 K. The sample with $x$ = 0.08 reveals different directions of SD lattice in adjacent domains, the typical angle between which is 27.8°. (k) Average dI/dV spectra taken within the above STM images.

**Figure 3:**

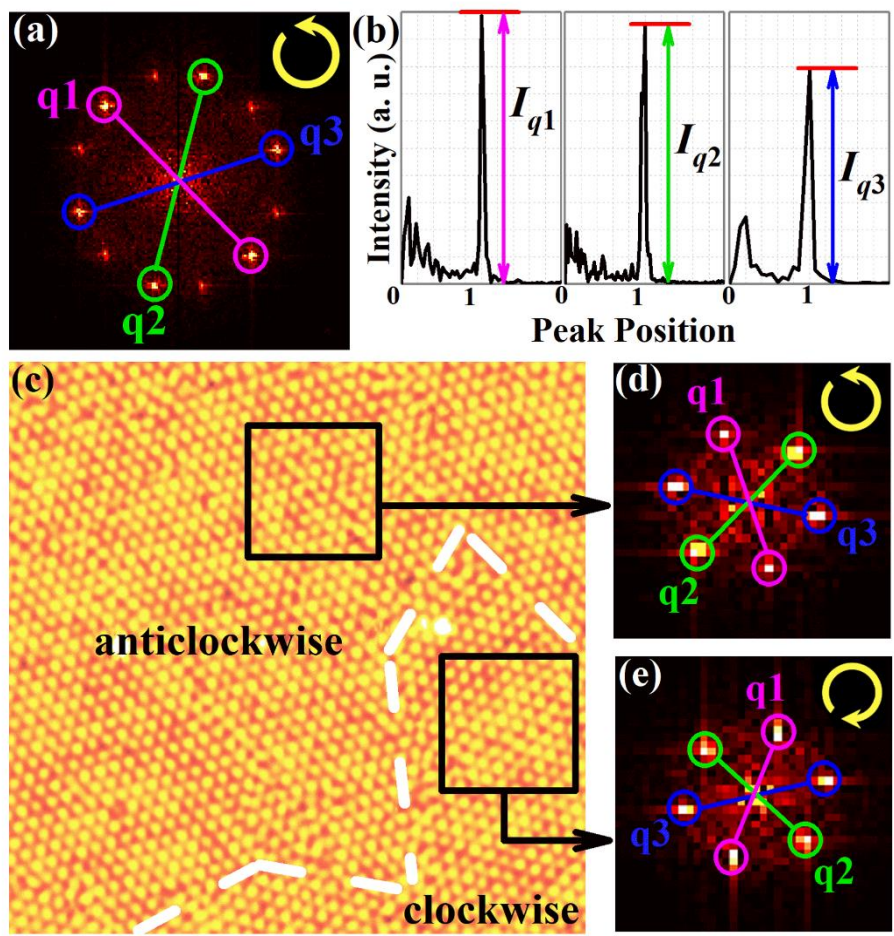

**Fig. 3 (color online):** (a) FT image of Fig. 2(g). (b) Line profiles along the **q1**, **q2**, **q3** wave vector. $I_{qi}$ is the intensity of the line profile along **qi**. (c) A 44 nm × 44 nm STM image for $x$ = 0.08. (d), (e) FT of two areas, each enclosed with a square in (c).

**Figure 4:**

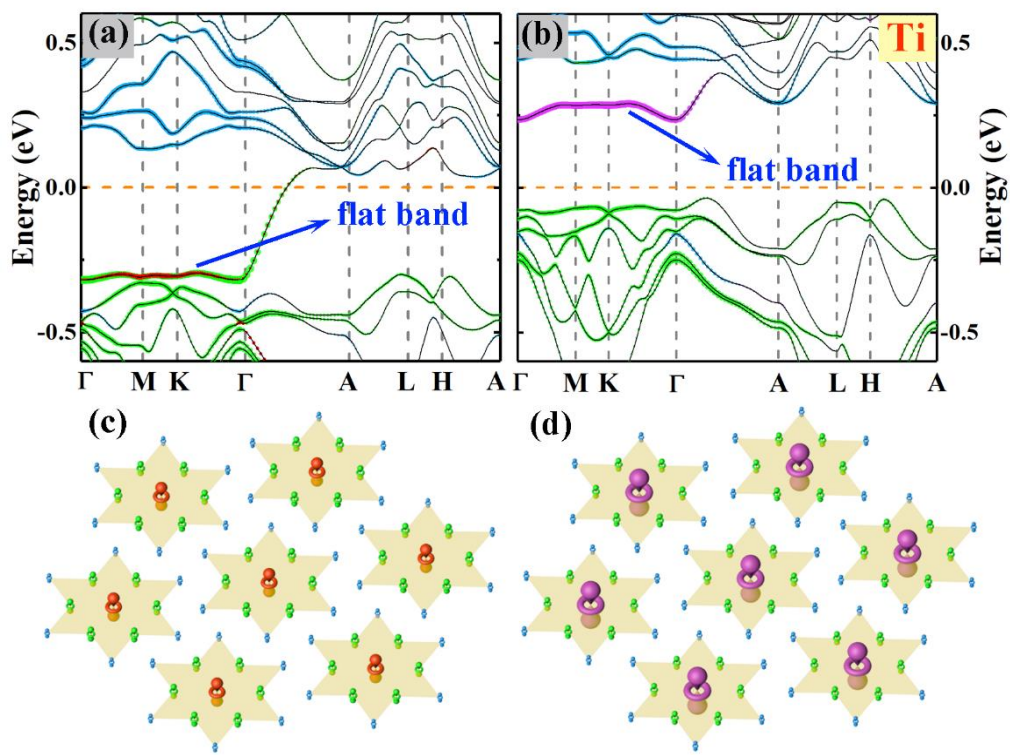

**Fig. 4 (color online):** Band structures of (a) $1T$-TaS$_2$ and (b) $1T$-Ti$_{1/13}$Ta$_{12/13}$S$_2$. The effective Hubbard $U_{eff}$ = 2.26 and 3.9 eV were added for the Ta and Ti, respectively. Red/purple, green and blue correspond to the flat bands of $d_{z^2}$ of Ta$_0$/Ti, Ta$_1$ and Ta$_2$, respectively. (c), (d) Schematic diagram of the orbital density wave order in the Ta layer. Each yellow star presents a cluster of thirteen Ta atoms, and the central atom is Ta/Ti in (c)/(d).